\documentclass[conference]{IEEEtran}
\IEEEoverridecommandlockouts
\usepackage{amsmath,amssymb,amsfonts}
\usepackage{algorithm}
\usepackage{algpseudocode}
\usepackage{graphicx}
\usepackage{textcomp}
\usepackage{xcolor}

\def\BibTeX{{\rm B\kern-.05em{\sc i\kern-.025em b}\kern-.08em
T\kern-.1667em\lower.7ex\hbox{E}\kern-.125emX}}

\usepackage[style=ieee, doi=false, url=false, isbn=false, giveninits=true, backend=biber, sorting=none, sortcites=true, date=year]{biblatex}
\begin{document}

\title{Timely Remote Estimation with Memory at the Receiver

\thanks{This work was supported in part by the NSF grant CNS-2239677.}
}

\author{\IEEEauthorblockN{Sirin Chakraborty}
\IEEEauthorblockA{\textit{Dept. of ECE} \\
\textit{Auburn University}\\
Auburn, AL, USA \\
szc0260@auburn.edu}
\and
\and
\IEEEauthorblockN{Yin Sun}
\IEEEauthorblockA{\textit{Dept. of ECE} \\
\textit{Auburn University}\\
Auburn, AL, USA \\
yzs0078@auburn.edu}
}
\maketitle

\thispagestyle{empty}
\begin{abstract}
In this study, we consider a remote estimation system that estimates a time-varying target based on sensor data transmitted over wireless channel. Due to transmission errors, some data packets fail to reach the receiver. To mitigate this, the receiver uses a buffer to store recently received data packets, which allows for more accurate estimation from the incomplete received data. Our research focuses on optimizing the transmission scheduling policy to minimize the estimation error, which is quantified as a function of the age of information vector associated with the buffered packets. Our results show that maintaining a buffer at the receiver results in better estimation performance for non-Markovian sources. 
\end{abstract}



\section{Introduction}
Timely status updates from sensors is a cornerstone for  networked intelligent systems that rely on live data to make estimations and real-time decisions. It enables these systems to provide timely estimations, leading to intelligent and proactive actions. For instance, autonomous vehicles depend on real-time state estimations to make safety-critical decisions, ensuring accident avoidance and smoother traffic integration. Similarly, remote healthcare systems use timely inference to monitor vital signs, allowing quick responses to health emergencies of remote patients. In industrial IoT systems, real-time fault detection ensures operational efficiency by addressing malfunctions before they escalate into significant problems. But due to data loss, transmission error, transmission delay the status updates are not always fresh, which significantly affects the accuracy of timely predictions. To evaluate the freshness of data updates, we use the \textit{Age of Information (AoI)} metric. Traditionally, AoI is defined as the time difference between the current time $t$ and the generation time $U(t)$ of the freshest packet delivered to the receiver, such that $\Delta(t)=t-U(t)$. A smaller AoI indicates the presence of more recent information at the receiver. 


\begin{figure}[t]
    \centering
    \includegraphics[scale=0.118]{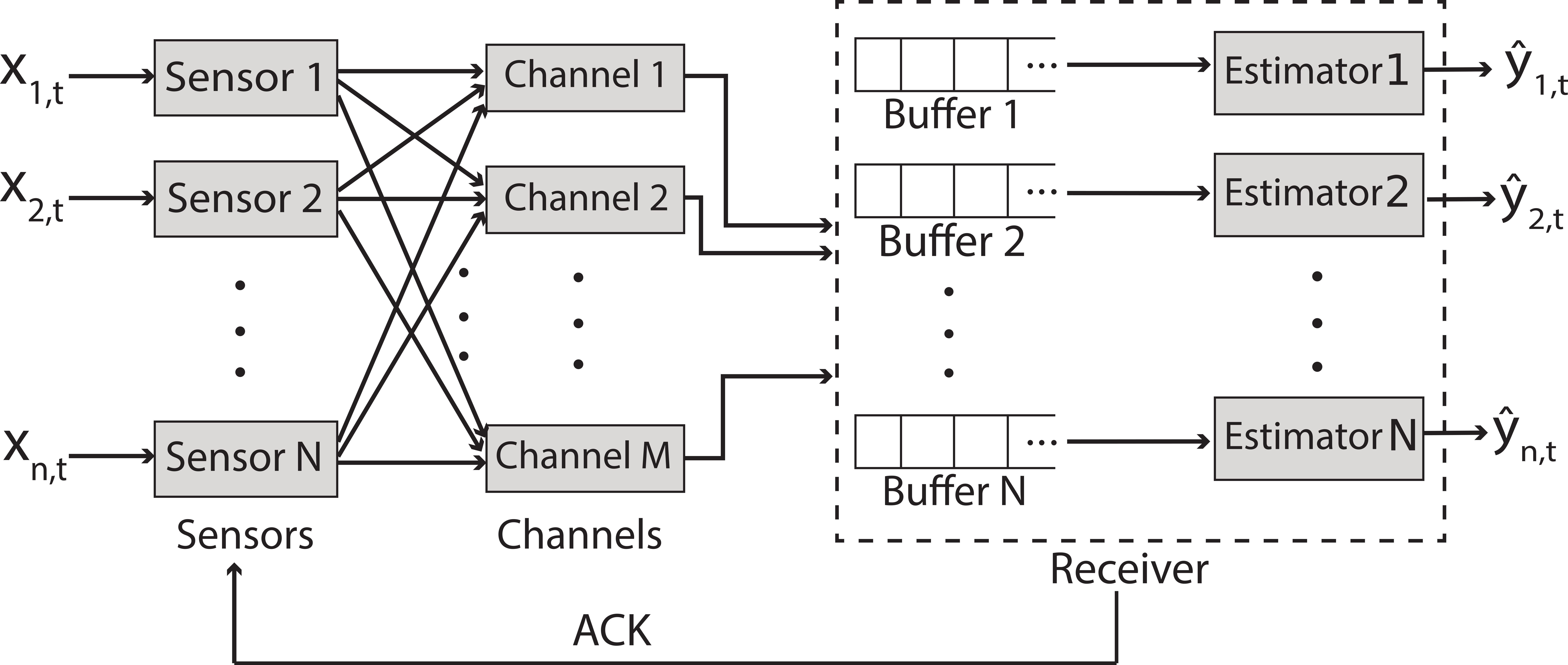}
    \caption{System Model}
    \label{sys_model}
\end{figure}

In this paper, we consider a discrete-time remote estimation system  for non-Markovian sources that consists of multiple sensors transmitting observations to a common receiver, as shown in Figure \ref{sys_model}. Due to channel sharing among sensors and potential transmission errors, the data at the receiver may become outdated. To address the challenges posed by data staleness and the non-Markovian nature of the sources, we propose a buffer-based remote estimation model. In this model, the receiver maintains a buffer of size $b$ for each sensor $n$, storing the 
$b$ most recently received data packets. After each successful packet delivery from sensor $n$, the newly received data packet is stored in the buffer, while the oldest packet is discarded to maintain the fixed buffer size $b$. This setup extends the conventional concept of AoI to an AoI vector. Specifically, let $\mathbf{\Delta}_n(t) = (\Delta_{n,1}(t),\Delta_{n,2}(t), \ldots,\Delta_{n,b}(t))$ denote the AoI vector of the buffered packets received from sensor $n$, where $\Delta_{n,i}(t)$ represents the age of the $i$-th most recently delivered packet from sensor $n$. Therefore, the first element $\Delta_{n,1}(t)$ corresponds to the AoI of the freshest packet, while the subsequent elements capture the AoI values of older packets stored in the buffer. When $b=1$, the model reduces to the conventional remote estimation framework used for Markovian sources; see, e.g., \cite{sun_2017_remote,Sun_2019_JCN,klugel_2019_aoi,ornee_2021_sampling, Ornee_2023_MobiHoc}.  
The technical contributions of this paper are summarized as follows:
\begin{itemize}
    \item We formulate a transmission scheduling problem for buffer-based remote estimation, aiming to minimize the average estimation error of the sensors. At each time $t$, the estimation error for sensor $n$ is modeled as a function of its AoI vector $\mathbf{\Delta}_n(t)$. This framework ensures accurate real-time estimation of non-Markovian sources, by dynamically prioritizing sensors for transmission based on their AoI vectors and their impact on reducing the overall system error.

    \item The transmission scheduling problem is cast as a Restless Multi-Armed Bandit (RMAB) and solved using Lagrangian relaxation and dual decomposition. This method decomposes the relaxed problem into smaller sub-problems, which can be efficiently solved using dynamic programming. Leveraging the solution to the relaxed problem, a Maximum Gain First (MGF) scheduling policy is developed to address the original RMAB problem. Notably, this method eliminates the need to satisfy the indexability condition, which is typically required for Whittle index-based solutions. 
    \item Our numerical results show that maintaining a buffer at the receiver results in better estimation for non-Markovian sources. To the best of our knowledge, this is the first work to study AoI-vector-based scheduling for remote estimation with a receiver-side buffer.

\end{itemize}

\section{Related Works}

The Age of Information (AoI) has become a key metric for quantifying data freshness in networked systems since its introduction in \cite{kaul_2012_real}. Early research focused on optimizing average and peak AoI within communication networks \cite{yates_2021_age,yates_2015_lazy,sun_2017_update}. More recent studies have applied AoI to enhance real-time applications, including remote inference \cite{shisher_2021_age,shisher_2022_does}, edge computing \cite{muhammad_2021_minimizing}, and control systems \cite{yates_2021_age,klugel_2019_aoi}, making it a central optimization tool. However, AoI focuses solely on timeliness and ignores the relevance of updates. To address this, complementary metrics such as Age of Incorrect Information (AoII) \cite{maatouk_2020_age}, Age of Synchronization (AoS) \cite{zhong_2018_two}, and Value of Information (VoI) \cite{soleymani_2024_consistency} have been developed. Additionally, \cite{guo_2019_credibility} introduces the concept of credibility in real-time wireless networks, focusing on minimizing the system-wide Loss-of-Credibility (LoC) by ensuring timely packet deliveries. These metrics extend AoI by incorporating aspects like estimation error, context dependency, and the performance benefits of transmitting specific updates.


Previous works have extensively applied Kalman filters in remote estimation systems to address packet loss, delays, and communication constraints. \cite{sinopoli_2004_kalman} examined the stability of Kalman filtering over packet-erasure channels, identifying conditions for bounded estimation error despite intermittent observations. \cite{schenato_2008_optimal} analyzed Kalman filtering under random delays and packet drops, demonstrating that stability thresholds can be independent of delays under certain conditions. However, other studies, such as \cite{sun_2017_remote} and \cite{ornee_2021_sampling}, highlight the interplay between AoI and remote estimation by designing optimal sampling strategies to minimize AoI-based estimation error. Building on this, our work incorporates the AoI as a vector into the estimation and scheduling process, enhancing accuracy under network constraints.

The optimization of both linear and non-linear AoI functions for multiple source scheduling can be framed as a restless multi-armed bandit (RMAB) problem, as explored in several works \cite{tripathi_2019_whittle,shisher_2022_does, chen_2021_scheduling,shisher_2024_timely,shisher_2023_learning,ornee_2023_context}. Whittle's pioneering work \cite{whittle_1988_restless} introduced an index-based policy to solve RMAB problems with binary actions. An extension of the Whittle index policy for handling multiple actions was presented in \cite{Hodge_2015_Ota}, though it requires satisfying a complex indexability condition. In addition to Whittle index-based policies, which rely on satisfying an indexability condition, non-indexable scheduling policies have also been explored in studies such as \cite{shisher_2023_learning,chen_2023_index,ornee_2023_context,chen_2021_scheduling}. Due to the complex buffer mechanism and channel fading conditions, Whittle index theory could not be applied to establish indexability for our multi-source scheduling problem. To address this, we adopt the ``Net-gain Maximization'' policy from our recent work \cite{shisher_2023_learning,ornee_2023_context} and rename it as the ``Maximum Gain First'' policy. This policy does not rely on the indexability condition.
It was also called the Optimal Lagrange Index policy in \cite{brown_2020_index}, LP-Index policy in \cite{gast_2023_linear}, and Gain Index policy in \cite{chen_2023_index}.

\section{Model and Formulation}

\subsection{System Model}\label{AA}
We consider the discrete-time remote estimation system depicted in Figure \ref{sys_model}, where $N$ sensors transmit status updates over $M$ shared wireless channels to a common receiver. At each time slot $t\in\{0,1,2,\ldots\}$, the receiver estimates a time-varying target $Y_{n,t} \in \mathcal{Y}$ using status update packets received from sensor $n$ up to time $t$. A scheduler decides which sources to select for data transmission. The scheduler's decision for sensor $n$ at time $t$ is represented by an indicator function
\begin{align}
    u_n(t)=\begin{cases}
        1, & \text{if decides to transmit},\\
        0, & \text{otherwise}.
    \end{cases}
\end{align}
In response to the scheduler's decision, each sensor $n$ submits a time-stamped status update packet $(X_{n,t},t)$ of its observation $X_{n,t}\in \mathcal{X}$ to a channel. However, due to transmission errors over wireless fading channels, some  packets are lost during the transmissions. This transmission successful events can be expressed using indicator functions $c_n(t)\in\{0,1\}$ that are i.i.d.~across time and sensors. The transmission successful probability is denoted by $p_n = \Pr\{c_{n}(t) = 1 \}$. The freshest information the receiver gets at time $t$ is $X_{n,t-\Delta_n(t)}$, which was generated $\Delta_n(t)$ time slots ago. This time difference $\Delta_n(t)$ is the AoI of sensor $n$. If an update is sent at beginning of time slot $t$, it gets delivered at beginning of time slot $t+1$, then the AoI evolution of the $n$-th source follows
\begin{align}\label{eq_AoI}
    \Delta_n (t+1)=
    \begin{cases}
        1, & \text{if $c_n(t)u_n(t)=1$},\\
        \Delta_n (t) + 1, & \text{otherwise}.
    \end{cases}
\end{align}
We assume that at a given time, each channel can serve only one of the $N$ sources and after every successful transmission, the receiver sends an error free acknowledgment back to the transmitter.


The receiver consists of $N$ buffers, each of a size $b$, connected to the $N$ estimators corresponding to the $N$ sensors. Each buffer $n$ can store at most $b$ packets previously received from sensor $n$. The content of buffer $n$ is 
$\{(X_{n,t-\Delta_{n,1}(t)}, \Delta_{n,1}(t)),(X_{n,t-\Delta_{n,2}(t)},\Delta_{n,2}(t)),\dots,$ $ 
(X_{n,t-\Delta_{n,b}(t)},\Delta_{n,b}(t))\}$, where the $i$-th packet consists of a stale observation $X_{n,t-\Delta_{n,i}(t)}$ from sensor $n$, generated $\Delta_{n,i}(t)$ time ago and its corresponding AoI $\Delta_{n,i}(t)$ such that $i\in\{1,2,\dots,b\}$. At time $t+1$, if the latest transmission is successful, i.e. $u_n(t)=1$, the buffer gets updated with the freshest packet $(X_{n,t},1)$ and discards the oldest data packet in order to maintain the fixed size of $b$. In contrast, if the transmission fails, i.e. $u_n(t)=0$, the signal values in the buffer contents stay the same, but their age values get one time slot older, mathematically represented as
$(X_{n,t-\Delta_{n,i}}(t),\Delta_{n,i}(t)+1)$. The transition of the AoI vector is expressed as \begin{align}
    &\mathbf{\Delta}_n(t+1)=\\ \nonumber
    &\begin{cases}
        (1,\Delta_{n,1}(t)+1,\ldots,\Delta_{n,b-1}(t)+1), ~~~\text{if } c_n(t) u_n(t)=1,\\
        (\Delta_{n,1}(t)+1,\Delta_{n,2}(t)+1,\ldots,\Delta_{n,b}(t)+1), ~\text{otherwise}.
    \end{cases}
\end{align}



Estimator $n$ takes the contents of buffer $n$ as a feature vector of length $b$ and generates an output $a=\phi_n(\mathbf{X}_{n,t-\mathbf{\Delta}_{n}(t)},\mathbf{\Delta}_n(t))\in \mathcal{A}$, where $\mathbf{X}_{n,t-\mathbf{\Delta}_n(t)}=(X_{n,t-\Delta_{n,1}(t)}, X_{n,t-\Delta_{n,2}(t)},\dots,
X_{n,t-\Delta_{n,b}(t)})$ is the feature vector of length $b$, $\mathbf{\Delta}_n(t)=(\Delta_{n,1}(t),\Delta_{n,2}(t),\dots,\Delta_{n,b}(t))$ is the age vector of the packets in the feature vector, and $\phi:(\mathcal{X}^b\times\mathbb{Z^+}^b)\mapsto \mathcal{A}$ is the estimation function. We consider an estimator whose estimation performance for sensor $n$ at time slot $t$ is measured by a loss function $L_n:\mathcal{Y}\times \mathcal{A}\mapsto \mathbb{R}$, where $L_n(y,a)$ indicates the incurred loss if the output $a$ is used for estimation when $Y_{n,t}=y$. The estimation problem is formulated as
\begin{align}
    &\mathrm{err}_n(\mathbf{\Delta}_n(t))\nonumber\\=&\min_{\phi\in\Phi}\mathbb{E}
\left[L_n(Y_{n,t},\phi_n(\mathbf{X}_{n,t-\mathbf{\Delta}_n(t)},\mathbf{\Delta}_n(t)))
|\mathbf{\Delta}_n(t)\right]\label{err_eq}.
\end{align}

The loss function $L_n$ is defined based on the objective of the remote estimation system \cite{shisher_2024_timely}. For instance, in a neural network designed for minimum mean-squared error estimation, the loss function is given by $L_n(y,\hat{y})=(y-\hat{y})^2$, where the action $a=\hat{y}$ represents the estimate of the target $Y_{n,t}=y$. In the case of softmax regression, which is a neural network-based approach for maximum likelihood classification, the action $a=Q_Y$ corresponds to a probability distribution over $Y_{n,t}$, and the loss function $L(y,Q_Y)=-\log(Q_Y(y))$ represents the negative log-likelihood of the target value $Y_{n,t}=y$.







\subsection{Scheduling Policy and Problem Formulation}
A scheduling policy is denoted by $\pi=(\pi_n)_{n=1}^N$, where $\pi_n=(u_n(0),u_n(1),\dots)$. Let $\Pi$ be the set of all possible causal scheduling policies, in which every decision $u_n(t)$ is made using the current and history information available at the scheduler. As our system consists of $M$ different channels, the condition $\sum_{n=1}^N u_n(t)\leq M$ is required to hold for all $t$.

Our goal is to find the optimal scheduling policy that minimizes the discounted sum of the expected estimation errors among all the $N$ sources over an infinite time-horizon. So the optimization problem can be formulated as
\begin{align}
\inf_{\pi\in\Pi}&\limsup_{T\to\infty} \sum_{n=1}^N \sum_{t=0}^{\infty}\mathbb{E}_\pi\left[\gamma^t \mathrm{err}_n(\mathbf{\Delta}_n(t))\right], \label{problem}\\
~\text{s.t.}~\!&\sum_{n=1}^{N} u_n (t) \leq M, u_n (t) \in \{0,1\}, t = 0,1, \ldots\label{constraint},
\end{align}
where $\gamma\in [0,1]$ is the discount factor and $\mathbb{E}_\pi[\cdot]$ is the expectation under policy $\pi$.

\section{Restless Multi-Armed Bandit Solution}
Problem \eqref{problem}-\eqref{constraint} is an RMAB problem, considering each source $n$ as an arm with $\mathbf{\Delta}_n(t)$ as the state. To solve RMAB problems, the Whittle index policy in \cite{whittle_1988_restless} is the most commonly used method. However, a key challenge in applying the Whittle index is satisfying the problem's indexability condition. Due to the complex buffer state transitions, proving indexibility is challenging in our problem. Therefore, we use a different algorithm, following \cite{shisher_2023_learning}, which does not require satisfying the indexability condition.

\subsection{Lagrangian Relaxation and Dual Decomposition}
To address \eqref{problem}-\eqref{constraint}, we first relax the per-time-slot channel constraints \eqref{constraint} into a discounted time-summation form, as expressed in constraint \eqref{relax_constraint} of the following relaxed problem:
\begin{align}
\inf_{\pi\in\Pi}&\limsup_{T\to\infty} \sum_{n=1}^N\sum_{t=0}^{\infty}\gamma^t\mathbb{E}_\pi\left[ \mathrm{err}_n(\mathbf{\Delta}_n(t))\right], \label{relax_problem}\\
 ~\text{s.t.}~\!&\limsup_{T\to\infty}{\sum_{n=1}^N \sum_{t=0}^{\infty}\gamma^t\mathbb{E}_{\pi}\left[u_n(t)\right]}\leq \frac{M}{1-\gamma}\label{relax_constraint}.
\end{align}
 To solve this relaxed problem, we apply the Lagrange dual decomposition method \cite{whittle_1988_restless}, using a Lagrange multiplier $\lambda\geq 0$. The dual problem is formulated as
\begin{equation}
    \lambda^*=\arg\max_{\lambda\geq 0}{q}(\lambda)\label{opt_dual},
\end{equation}
where the dual function ${q}(\lambda)$ is defined as
\begin{align}\label{lag_prob}
    &{q}(\lambda)=\\ \nonumber
&\inf_{\pi\in\Pi}\limsup_{T\to\infty} \sum_{n=1}^N\sum_{t=0}^{\infty}\gamma^t\mathbb{E}_\pi\left[ \mathrm{err}_n(\mathbf{\Delta}_n(t))
+\lambda u_n(t)\right]-\frac{\lambda M}{1-\gamma}. 
\end{align}
Given $\lambda$, problem \eqref{lag_prob} can be decomposed into $N$ sub-problems, and the sub-problem for sensor $n$ is formulated as
\begin{equation}
\inf_{\pi_n\in\Pi_n}\limsup_{T\to\infty}\sum_{t=0}^{\infty}\gamma^t\mathbb{E}_{\pi_n}\left[\mathrm{err}_n(\mathbf{\Delta}_n(t))+\lambda u_n(t)\right], \label{sub_problem}
\end{equation}
where $\Pi_n$ is the set of all causal scheduling policies $\pi_n=(u_n(0),u_n(1),\dots)$ of sensor $n$. 
The Lagrange multiplier $\lambda\geq0$ can be considered as a transmission cost, which sensor $n$ has to pay for using a channel resource.

\subsection{Optimal Solution for the  Relaxed Problem \eqref{relax_problem}-\eqref{relax_constraint}}
The sub-problem \eqref{sub_problem} for each sensor $n$ is a discounted infinite-horizon MDP. Suppose that the current state $\mathbf{\Delta}_n(t)$ and action $u_n(t)$ are denoted as $s$ and $a$, respectively. Then, the Bellman optimality equation for the MDP \eqref{sub_problem} is

\begin{equation}\label{val_func}
    V^*_{n,\lambda}(s)=\min_{a\in\{0,1\}} Q^*_{n,\lambda}(s,a),
\end{equation}
\begin{equation}\label{val_func1}
    Q^*_{n,\lambda}(s,a)=\mathrm{err}_n(s)+\lambda a+\gamma \sum_{s'} p_{ss'}^aV_{n,\lambda}^*(s'),
\end{equation}
where $V^*_{n,\lambda}(\cdot)$ and $Q^*_{n,\lambda}(\cdot,\cdot)$ are the optimal value function and the optimal action-value function of the MDP \eqref{sub_problem}, respectively. The action-value function $Q^*_{n,\lambda}(s,a)$ is also known as the Q-function or Q-value. Given any $\lambda\geq0$, we use a dynamic programming algorithm to solve \eqref{val_func}-\eqref{val_func1}.

Next, we use the stochastic sub-gradient ascent method to solve \eqref{opt_dual} and obtain the optimal dual variable $\lambda^*$. At each iteration $k$, the dual variable is updated as follows:
\begin{equation}
    \lambda(k+1)=\lambda(k)+\frac{\beta}{k} \bigg[\sum_{n=1}^{N}\sum_{t=0}^\infty \gamma^t u^*_{n,\lambda(k)}(t)-\frac{M}{1-\gamma}\bigg],\label{stoch_sub_grad}
\end{equation}
where $\beta>0$ is a constant, and $u^*_{n,\lambda(k)}(t)$ represents whether sensor $n$ is scheduled at time slot $t$ in the optimal solution to the sub-problem \eqref{sub_problem} when $\lambda = \lambda(k)$.

\subsection{Maximum Gain First Policy for Problem \eqref{problem}-\eqref{constraint}}
Now we present the Maximum Gain First (MGF) scheduling policy, which is a feasible solution to the original problem \eqref{problem}-\eqref{constraint}. 
Using the optimal dual variable $\lambda^*$ of the relaxed problem \eqref{relax_problem}-\eqref{relax_constraint}, the gain $\alpha_{n,\lambda^*}(\mathbf{\Delta}_n(t))$ of scheduling sensor $n$ at time slot $t$ is defined as the difference of Q-values between not transmitting and transmitting \cite{shisher_2023_learning,ornee_2023_context}:
\begin{equation}\label{gain}
    \alpha_{n,\lambda^*}(\mathbf{\Delta}_n(t))=Q^*_{n,\lambda^*}(\mathbf{\Delta}_n(t),0)-Q^*_{n,\lambda^*}(\mathbf{\Delta}_n(t),1).
\end{equation}
At each time slot $t$, the MGF policy selects no more than  $M$ sensors that have the highest non-negative gains $\alpha_{n,\lambda^*}(\mathbf{\Delta}_n(t))$, as described in Algorithm \ref{alg1}.  We note that the MGF policy is also the solution to the following problem: 
\begin{align}
    \max_{u_n(t),n=1,\ldots,N} &\sum_{n=1}^N \alpha_{n,\lambda^*}(\mathbf{\Delta}_n(t))u_n(t)\\
    \text{s.t.}~~~~~~ &\sum_{n=1}^N u_n(t)\leq M,~u_n (t) \in \{0,1\}.
\end{align}

\begin{algorithm}[t]
\caption{Maximum Gain First Scheduling Policy}
\begin{algorithmic}[1]
\State \textbf{Input:} Optimal Lagrange multiplier $\lambda^*$ obtained by solving \eqref{relax_problem}-\eqref{relax_constraint} using \eqref{val_func}-\eqref{stoch_sub_grad}.
\For{each time step \( t =1,2,\ldots \)}
    \For{each sensor \( n = 1,2,\ldots, N \)}
        \State \text{Update state} \( \mathbf{\Delta}_n(t) \) based on \eqref{eq_AoI}.
        
        \State \text{Update the gain} \( \alpha_{n,\lambda^*}(\mathbf{\Delta}_n(t)) \) based on \eqref{gain}.
        
            \EndFor
    
    \State Select at most $M$ sensors with the highest non-negative gains \( \alpha_{n,\lambda^*}(\mathbf{\Delta}_n(t)) \).

                \EndFor

\end{algorithmic} \label{alg1}
\end{algorithm}

\section{Evaluation}

\begin{figure}[t]
    \centering
    \includegraphics[scale=0.342]{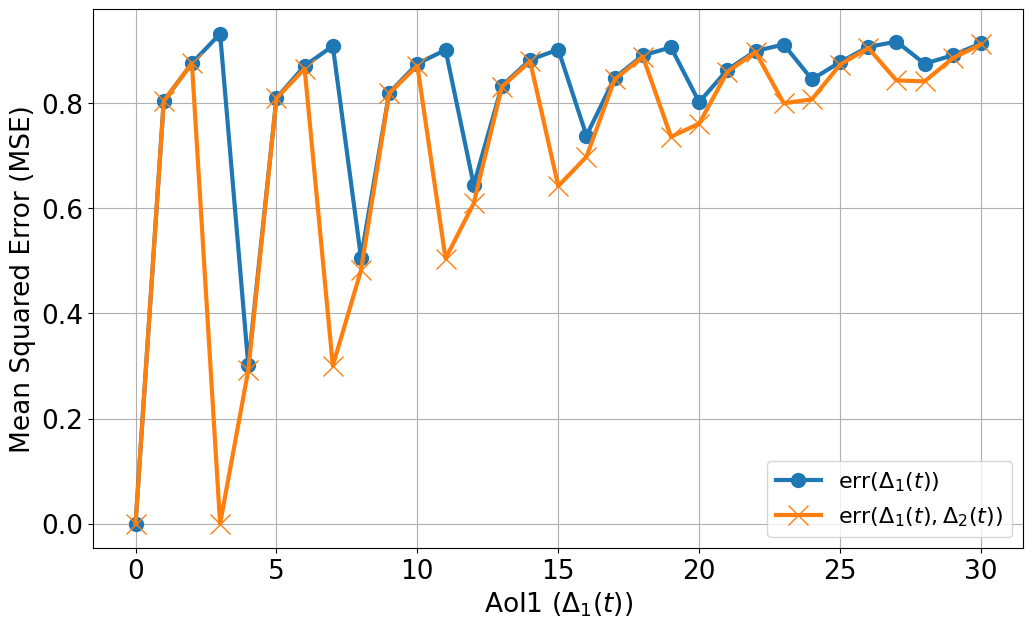}
    \caption{AoI $\Delta_1(t)$ vs estimation error with fixed $\Delta_2(t)=3$.}
    \label{aoiVMSE}
\end{figure}
We evaluate the performance of buffer-based remote estimation using the following fourth-order autoregressive system:
\begin{equation}
  X_{n,t}=0.1X_{n,t-1}+0.8X_{n,t-4}+W_{n,t},
\end{equation}
where $W_{n,t}\in\mathbb{R}$ represents i.i.d. Gaussian noise with zero mean and unit variance. The receiver uses a Kalman filter to estimate the current system state $X_{n,t}$, based on previously received data packets that are stored in the buffer. 
The mean squared error (MSE) of the Kalman filter is a function of AoI. Figure \ref{aoiVMSE} depicts the MSE as a function of AoI under two scenarios: (i) buffer size $b=1$, where the MSE is represented as \(\text{err}(\Delta_1(t))\), and (ii) buffer size $b=2$, where the MSE is represented as \(\text{err}(\Delta_1(t), \Delta_2(t))\). In both cases, the MSE is found to be a non-monotonic function of AoI, consistent with the results reported in \cite{shisher_2024_timely}. One can observe that storing two data packets in the buffer reduces the MSE compared to the single-packet case in traditional remote estimation framework.

\begin{figure}[t]
    \centering
    \includegraphics[width=0.44\textwidth]{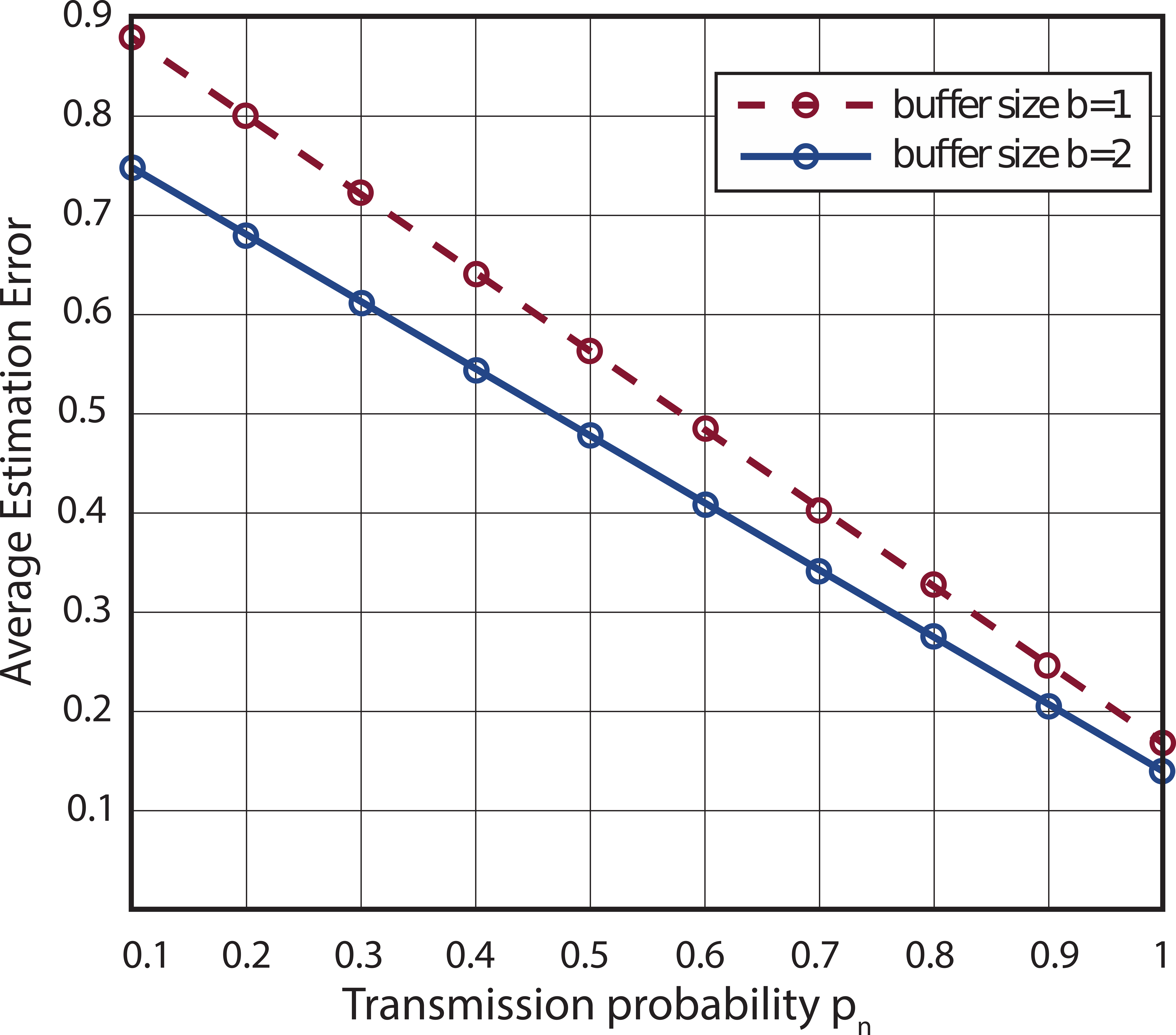}
    \caption{Average estimation error vs transmission probability $p_n$.}
    \label{errorVp}
\end{figure}

We evaluate the performance of our proposed scheduling algorithm in minimizing the overall estimation error. The system consists of two sensors ($N=2$) and one communication channel ($M=1$). To assess the impact of buffer size at the receiver, we analyze two scenarios: (i) a buffer that holds a single data packet for each sensor ($b=1$) and (ii) a buffer storing two data packets for each sensor ($b=2$). Figure \ref{errorVp} presents the average estimation error as a function of the transmission probability $p_n$, assuming equal transmission success probabilities for both sensors, i.e.,  $p_1=p_2$. The results demonstrate that, utilizing a buffer of size 2  reduces the optimized average estimation error compared to a buffer of size 1.

\section{Conclusion}
In this paper, we design a remote estimation system with buffers at the receiver storing historically received data. Our results show that maintaining a buffer of multiple packets reduces the estimation error for non-Markovian sources.
\section{Acknowledgement}
The authors appreciate the encouragement from I-Hong Hou, Roy D. Yates, and Shaolei Ren to consider incorporating a buffer at the receiver.
\printbibliography

@inproceedings{shisher_2021_age,
  title={The age of correlated features in supervised learning based forecasting},
  author={Shisher, Md Kamran Chowdhury and Qin, Heyang and Yang, Lei and Yan, Feng and Sun, Yin},
  booktitle={IEEE INFOCOM AoI Workshop},
  pages={1--8},
  year={2021},
}

@inproceedings{shisher_2022_does,
  title={How does data freshness affect real-time supervised learning?},
  author={Shisher, Md Kamran Chowdhury and Sun, Yin},
  booktitle={ACM MobiHoc},
  pages={31--40},
  year={2022}
}

@inproceedings{kaul_2012_real,
  title={Real-time status: How often should one update?},
  author={Kaul, Sanjit and Yates, Roy and Gruteser, Marco},
  booktitle={2012 Proceedings IEEE INFOCOM},
  pages={2731--2735},
  year={2012}
}

@article{muhammad_2021_minimizing,
  title={Minimizing age of information in multiaccess-edge-computing-assisted IoT networks},
  author={Muhammad, Ali and Sorkhoh, Ibrahim and Samir, Moataz and Ebrahimi, Dariush and Assi, Chadi},
  journal={IEEE Internet of Things Journal},
  volume={9},
  number={15},
  pages={13052--13066},
  year={2021},
  publisher={IEEE}
}

@article{yates_2021_age,
  title={Age of information: An introduction and survey},
  author={Yates, Roy D and Sun, Yin and Brown, D Richard and Kaul, Sanjit K and Modiano, Eytan and Ulukus, Sennur},
  journal={IEEE Journal on Selected Areas in Communications},
  volume={39},
  number={5},
  pages={1183--1210},
  year={2021}
}

@inproceedings{klugel_2019_aoi,
  title={{AoI}-penalty minimization for networked control systems with packet loss},
  author={Kl{\"u}gel, Markus and Mamduhi, Mohammad H and Hirche, Sandra and Kellerer, Wolfgang},
  booktitle={IEEE INFOCOM AoI Workshop},
  pages={189--196},
  year={2019}
}

@inproceedings{yates_2015_lazy,
  title={Lazy is timely: Status updates by an energy harvesting source},
  author={Yates, Roy D},
  booktitle={IEEE ISIT},
  pages={3008--3012},
  year={2015}
}

@article{sun_2017_update,
  title={Update or wait: How to keep your data fresh},
  author={Sun, Yin and Uysal-Biyikoglu, Elif and Yates, Roy D and Koksal, C Emre and Shroff, Ness B},
  journal={IEEE Transactions on Information Theory},
  volume={63},
  number={11},
  pages={7492--7508},
  year={2017}
}

@article{maatouk_2020_age,
  title={The age of incorrect information: A new performance metric for status updates},
  author={Maatouk, Ali and Kriouile, Saad and Assaad, Mohamad and Ephremides, Anthony},
  journal={IEEE/ACM Transactions on Networking},
  volume={28},
  number={5},
  pages={2215--2228},
  year={2020}
}

@inproceedings{zhong_2018_two,
  title={Two freshness metrics for local cache refresh},
  author={Zhong, Jing and Yates, Roy D and Soljanin, Emina},
  booktitle={IEEE ISIT},
  pages={1924--1928},
  year={2018},
}

@article{soleymani_2024_consistency,
  title={Consistency of value of information: Effects of packet loss and time delay in networked control systems tasks},
  author={Soleymani, Touraj and Baras, John S and Wang, Siyi and Hirche, Sandra and Johansson, Karl H},
  journal={arXiv preprint arXiv:2403.11932},
  year={2024}
}

@article{sinopoli_2004_kalman,
  title={Kalman filtering with intermittent observations},
  author={Sinopoli, Bruno and Schenato, Luca and Franceschetti, Massimo and Poolla, Kameshwar and Jordan, Michael I and Sastry, Shankar S},
  journal={IEEE transactions on Automatic Control},
  volume={49},
  number={9},
  pages={1453--1464},
  year={2004}
}

@article{schenato_2008_optimal,
  title={Optimal estimation in networked control systems subject to random delay and packet drop},
  author={Schenato, Luca},
  journal={IEEE transactions on automatic control},
  volume={53},
  number={5},
  pages={1311--1317},
  year={2008},
}

@inproceedings{sun_2017_remote,
  title={Remote estimation of the Wiener process over a channel with random delay},
  author={Sun, Yin and Polyanskiy, Yury and Uysal-Biyikoglu, Elif},
  booktitle={IEEE ISIT},
  pages={321--325},
  year={2017}
}

@article{ornee_2021_sampling,
  title={Sampling and remote estimation for the {O}rnstein-{U}hlenbeck process through queues: Age of information and beyond},
  author={Ornee, Tasmeen Zaman and Sun, Yin},
  journal={IEEE/ACM Transactions on Networking},
  volume={29},
  number={5},
  pages={1962--1975},
  year={2021}
}

@inproceedings{tripathi_2019_whittle,
  title={A {W}hittle index approach to minimizing functions of age of information},
  author={Tripathi, Vishrant and Modiano, Eytan},
  booktitle={2019 57th Annual Allerton Conference on Communication, Control, and Computing (Allerton)},
  pages={1160--1167},
  year={2019}
}

@article{chen_2021_scheduling,
  title={Scheduling to minimize age of incorrect information with imperfect channel state information},
  author={Chen, Yutao and Ephremides, Anthony},
  journal={Entropy},
  volume={23},
  number={12},
  pages={1572},
  year={2021}
}

@article{whittle_1988_restless,
  title={Restless bandits: Activity allocation in a changing world},
  author={Whittle, Peter},
  journal={Journal of applied probability},
  volume={25},
  number={A},
  pages={287--298},
  year={1988},
  publisher={Cambridge University Press}
}

@Article{Hodge_2015_Ota,
  author  = {Hodge, D. J. and Glazebrook, K. D.},
  journal = {Adv. Appl. Probab.},
  title   = {On the asymptotic optimality of greedy index heuristics for multi-action restless bandits},
  year    = {2015},
  number  = {3},
  pages   = {652–667},
  volume  = {47},
  doi     = {10.1239/aap/1444308876},
}

@article{shisher_2023_learning,
  title={Learning and communications co-design for remote inference systems: Feature length selection and transmission scheduling},
  author={Shisher, Md Kamran Chowdhury and Ji, Bo and Hou, I-Hong and Sun, Yin},
  journal={IEEE Journal on Selected Areas in Information Theory},
year={2023},
  volume={4},
  number={},
  pages={524-538}
}

@article{chen_2023_index,
  title={An index policy for minimizing the uncertainty-of-information of {M}arkov sources},
  author={Chen, Gongpu and Liew, Soung Chang},
  journal={IEEE Transactions on Information Theory},
  year={2024},
  volume={70},
  number={1},
  pages={698-721}
}

@article{brown_2020_index,
  title={Index policies and performance bounds for dynamic selection problems},
  author={Brown, David B and Smith, James E},
  journal={Management Science},
  volume={66},
  number={7},
  pages={3029--3050},
  year={2020}
}

@article{gast_2023_linear,
  title={Linear program-based policies for restless bandits: Necessary and sufficient conditions for (exponentially fast) asymptotic optimality},
  author={Gast, Nicolas and Gaujal, Bruno and Yan, Chen},
  journal={Mathematics of Operations Research},
  year={2023}
}

@inproceedings{ornee_2023_context,
  title={Context-aware Status Updating: Wireless Scheduling for Maximizing Situational Awareness in Safety-critical Systems},
  author={Ornee, Tasmeen Zaman and Shisher, Md Kamran Chowdhury and Kam, Clement and Sun, Yin},
  booktitle={IEEE MILCOM 2023},
  pages={194--200},
  year={2023}
}

@article{shisher_2024_timely,
  title={Timely communications for remote inference},
  author={Shisher, Md Kamran Chowdhury and Sun, Yin and Hou, I-Hong},
  journal={IEEE/ACM Transactions on Networking},
  year={2024},
  volume={32},
  number={5},
  pages={3824-3839}
}

@inproceedings{guo_2019_credibility,
  title={On the credibility of information flows in real-time wireless networks},
  author={Guo, Daojing and Hou, I-Hong},
  booktitle={2019 International Symposium on Modeling and Optimization in Mobile, Ad Hoc, and Wireless Networks (WiOPT)},
  pages={1--8},
  year={2019}
}

@ARTICLE{Sun_2019_JCN,
  author={Sun, Yin and Cyr, Benjamin},
  journal={Journal of Communications and Networks}, 
  title={Sampling for data freshness optimization: Non-linear age functions}, 
  year={2019},
  volume={21},
  number={3},
  pages={204-219}
}

@inproceedings{Ornee_2023_MobiHoc, 
author = {Ornee, Tasmeen Zaman and Sun, Yin}, 
title = {A {W}hittle Index Policy for the Remote Estimation of Multiple Continuous {G}auss-{M}arkov Processes over Parallel Channels}, 
year = {2023}, 
booktitle = {ACM MobiHoc}, 
pages = {91–100}}

\end{document}